\documentclass[epsf,aps,preprintnumbers,amsmath,amssymb,11pt]{revtex4}

\usepackage{epsf}

\usepackage{graphicx}

\usepackage{bm}

\begin {document}

%%%%%%%%%%%%%%%%%%%%%%%%%%%%%%%%%%%%%%%%%%%%%%%%%%%%%%%%%%%%%%%%%%%%%%%%%%%%%%%

%\def\Re{\operatorname{Re}}

\def\CP{CP^{N-1}}
\def\be{\begin{equation}}
\def\ee{\end{equation}}
\def\bea{\begin{eqnarray*}}
\def\eea{\end{eqnarray*}}

%%%%%%%%%%%%%%%%%%%%%%%%%%%%%%%%%%%%%%%%%%%%%%%%%%%%%%%%%%%%%%%%%%%%%%%%%%%%%%%

%\preprint {UW/PT 03--01}

\title
    {
    Coherent Topological Charge Structure in $CP^{N-1}$ Models and QCD 
    }

\author {Saeed Ahmad$^1$, Jonathan T.\ Lenaghan$^2$, and H. B. Thacker$^1$}
\affiliation
    {%
 $^1$. Department of Physics,
    University of Virginia,
    P.O. Box 400714
    Charlottesville, VA 22901-4714\\
 $^2$ 
     American Physical Society, One Research Road,
Box 9000, Ridge, NY 11961-9000, USA   }%

\date{today}

\begin {abstract}%
    {%
In an effort to clarify the significance of the recent observation of long-range topological charge
coherence in QCD gauge configurations, we study the local topological charge distributions in two-dimensional
$CP^{N-1}$ sigma models, using the overlap Dirac operator to construct the lattice topological charge. 
We find long-range sign 
coherence of topological charge along extended one-dimensional structures in two-dimensional spacetime.  
We discuss the connection between the long range topological structure found in $CP^{N-1}$  
and the observed sign coherence along three-dimensional 
sheets in four-dimensional QCD gauge configurations.  In both   
cases, coherent regions of topological charge form along membrane-like
surfaces of codimension one. We show that the Monte Carlo results, for both two-dimensional and four-dimensional 
gauge theory, support a view of topological charge fluctuations
suggested by Luscher and Witten. In this framework, the observed membranes are associated with
boundaries between
``k-vacua,'' characterized by an effective local value of $\theta$ which jumps by $\pm 2\pi$ 
across the boundary. }% 
\end {abstract}

\maketitle
\thispagestyle {empty}

 %%%%%%%%%%%%%%%%%%%%%%%%%%%%%%%%%%%%%%%%%%%%%%%%%%%%%%%%%%%%%%%%%%%%%%%%%%%%%%%

\section {Introduction}
\label{sec:intro}

Lattice studies of topological susceptibility 
and of the flavor singlet pseudoscalar hairpin correlator 
in QCD \cite{Bardeen00} have confirmed the role of topological charge and 
the axial anomaly in generating a large mass for the $\eta'$ meson. 
Nevertheless, the local structure of the topological charge fluctuations responsible 
for these effects is much less well-understood. The interrelation between
topological charge fluctuations, chiral symmetry breaking, and confinement is also obscure. 
Both confinement and $\chi SB$ reflect the long range
structure of the QCD vacuum, so it is natural to ask whether topological charge
fluctuations exhibit any kind of coherent long range structure. Of course,
conventional long range order is ruled out by spectral considerations, since the pure glue
theory has a large mass gap.
The topological charge correlation length is determined by the lightest pseudoscalar
glueball mass, which lattice calculations have shown to be well above 2 GeV \cite{Bali01}.
Furthermore, as first pointed out by Seiler and Stamatescu \cite{Seiler},
in the continuum limit, the Euclidean topological charge two-point correlator 
$G(x)=\langle q(x)q(0)\rangle$ is required to be negative
for all nonzero separation $|x|\neq 0$ \cite{Vicari}. This is a consequence of reflection positivity, from which it follows that {\it real, propagating intermediate states (glueballs) give
a negative contribution to the Euclidean correlator}.
The only positive contribution to the topological susceptibility
(as determined from the integrated correlator) comes from a positive
contact term at $x=0$ \cite{Seiler}. In spite of the absence of conventional long-range 
order in pure-glue QCD, there are reasons to suspect that a more subtle kind of long range
order is present. Specifically, it is clear that the essential properties of the gauge fields
responsible for $\chi SB$ should be manifest in the pure glue theory, even without internal quark loops.
This becomes apparent, for example, if we characterize the glue field by the eigenvalues of the
associated Dirac operator. Even quenched gauge configurations will support Goldstone boson propagation and
exhibit a chiral condensate in the form of a finite density of near-zero Dirac eigenmodes (up to
quenched chiral log corrections). Thus, the long-range order responsible for $\chi SB$ is built into
the glue field, in spite of the fact that the pure glue theory has a very large mass gap.

Until recently, the ability to study topological charge structure numerically in lattice QCD has been 
hindered by the lack of a well-behaved lattice discretization of the local topological charge
density. Ultralocal operators constructed directly from the gauge links are found to exhibit a large  
amount of short-range noise which obscures any possible long-range coherence. 
The emergence of exactly chiral lattice Dirac operators and associated Ginsparg-Wilson relations has provided a
new and vastly improved calculational approach to topological charge studies. In this approach, the local
topological charge density,
\begin{equation}
q(x) = \frac{g^2}{48\pi^2} Tr F\tilde{F}
\end{equation}
is calculated not from the link fields, but from the exactly chiral lattice Dirac
operator \cite{Hasenfratz98} 
\begin{equation}
\label{eq:TC}
q(x) = \frac{1}{2}Tr\gamma^5D
\end{equation}
where $D$ satisfies GW relations 
\begin{equation}
\label{eq:GW}
 \{\gamma_5, D\} = \bar{a}\times D\gamma_5D 
\end{equation}
and $\bar{a}\propto$ lattice spacing.
This construction of the lattice topological charge density 
arises in the calculation of the axial $U(1)$ anomaly on the lattice \cite{Hasenfratz98}.
Unlike the topological charge operator constructed from the gauge field \cite{Luscher82}, the 
operator (\ref{eq:TC}) turns out to reveal the coherent structure of the topological charge distribution in Monte
Carlo configurations without any need for smoothing or cooling procedures. 

Using the overlap construction of topological charge density, a recent study of pure-glue QCD gauge configurations 
has provided striking evidence for long range topological charge structure in the form of extended 
three-dimensional sheets in four-dimensional spacetime \cite{Horvath03_struct}. A typical gauge configuration (in a periodic box of 
about $(1.5 fm)^4$) contains two sheets of opposite charge which are everywhere thin and close together, forming
a crumpled or folded dipole layer which occupies a large fraction of space-time. A more detailed study of the 
charge distribution within these sheets \cite{Horvath05_global} has shown that the membrane-like structures are inherently
global, in the sense that the topological charge distribution is not concentrated in localized lumps but rather is
distributed throughout the coherent sheets. By comparing results for three different lattice spacings, it was
also found that the thickness of the sheets scales to zero in the continuum limit, and that the presence of these sheets 
plays a central role in the appearance of the expected positive contact term in the topological charge
correlator. The alternating-sign layered arrangement of the coherent sheets give rise to the negative 
short-distance power-law behavior expected from the operator product  expansion \cite{Horvath05_corr}.

In this paper, we study the distribution of topological charge in the $\CP$ model using the overlap Dirac operator.
We find that the topological charge distributions calculated by the overlap method 
in Monte Carlo generated $\CP$ configurations exhibit long range structure which is quite
 analogous to that observed in four-dimensional QCD. In $\CP$, the structures are effectively one-dimensional regions
of coherence in two-dimensional spacetime. We note that in both cases the coherent regions have the dimensionality corresponding 
to the world volume of a domain wall (i.e. codimension 1), a fact that  will play a central role in 
our theoretical interpretation of the observed structure. 

\section{Topological Charge in Asymptotically Free Gauge Theories}

\subsection{Topological Susceptibility and Long Range Order in Chern-Simons Currents}
  The possibility that the vacuum of pure-glue QCD possesses a ``secret long-range order'' associated with topological
charge was first explored by Luscher \cite{Luscher78}, who pointed out that if the topological susceptiblity
$\chi_t$ is nonzero, it implies the presence of a zero-mass pole in the correlator of two Chern-Simons
currents. Let us define the {\it abelian} 3-index Chern-Simons tensor
\begin{equation}
\label{eq:CStensor}
A_{\mu\nu\rho} = -Tr\left(B_{\mu}B_{\nu}B_{\rho}+\frac{3}{2}B_{[\mu}\partial_{\nu}B_{\rho]}\right)
\end{equation}
where $B_{\mu}$ is the Yang-Mills gauge potential. We consider the Chern-Simons current that is dual to this tensor,
\begin{equation}
j_{\mu}^{CS} = \epsilon_{\mu\nu\rho\sigma}A_{\nu\rho\sigma}\,\,.
\end{equation}
Although $j_{\mu}^{CS}$ is not gauge invariant, its divergence is the gauge invariant topological charge density
\begin{equation}
\label{eq:csdiv}
\partial_{\mu}j_{\mu}^{CS} = Tr F\tilde{F} = 32\pi^2 q(x) \,\,.
\end{equation}
Choosing a covariant gauge, $\partial_{\mu}A_{\mu\nu\rho}=0$, the correlator of two Chern-Simons currents
has the form
\begin{equation}
\label{eq:cscorr}
\langle j_{\mu}^{CS}(x)j_{\nu}^{CS}(0)\rangle = \int \frac{d^4p}{(2\pi)^4}\; e^{-ip\cdot x}\; \frac{p_{\mu}p_{\nu}}{p^2} G(p^2) \,\,.
\end{equation}
From (\ref{eq:csdiv}) we see that $G(p^2)$ must have a $p^2=0$ pole whose residue is the topological susceptibility,
\begin{equation}
G(p^2) \sim \frac{\chi_t}{p^2} \,\,.
\end{equation}
Of course, this pole does not imply the existence of a physical massless particle, because the Chern-Simons
current is not gauge invariant.
The gauge invariant topological charge correlator $\langle q(x)q(0)\rangle$ has no pole and
remains short range. Note that the $1/p^2$ pole in $G(p^2)$ gives rise to a contact term
in the topological charge correlator.  

To clarify the nature of the long-range order in 4D Yang-Mills theory, Luscher \cite{Luscher78} drew on the analogy
with 2-dimensional $CP^{N-1}$ sigma models. 
The continuum action for the $\CP$ model is 
\be
S = \beta N \int d^2x \left( D_\mu {\bf z} \right)^\dagger \cdot D_\mu {\bf z} \,\, ,
\ee
where ${\bf z}$ is an $N$-component complex scalar field subject to the 
constraint ${\bf z}^\dagger \cdot {\bf z} = 1$, and the covariant derivative is 
\be
D_\mu = \partial_\mu + i A_\mu \,\, .
\ee  
Here $A_{\mu}$ is a $U(1)$ gauge field, but it is an auxiliary field with no kinetic $F_{\mu\nu}^2$ term.
It's equation of motion sets it equal to the flavor-singlet current,
\be
\label{eq:London}
A_{\mu} = J_{\mu}
\ee
where

\begin{equation}
\label{eq:current}
J_{\mu} = \frac{1}{2}i\left({\bf z}^{\dagger}\partial_{\mu}{\bf z} -(\partial_{\mu}{\bf z})^{\dagger}{\bf z} \right) \,\,.
\end{equation} 
The $A_{\mu}$ field can be integrated out to give a theory of self-interacting $z$-particles. On the other hand, 
if we integrate out the $z$'s, the effective low energy
Lagrangian for the gauge field includes a dynamically generated kinetic term which arises from
closed $z$-loops. This dynamically generated $F_{\mu\nu}^2$ term gives rise to a confining potential 
between test $U(1)$ charges, and is also the origin of the $p^2=0$ pole in the Chern-Simons current correlator
and hence of the nonzero topological susceptibility.
The continuum topological charge density operator is defined as 
\be
q(x) \equiv \frac{1}{2\pi}\epsilon_{\mu\nu}\partial_{\mu}A_{\nu}\,\, .
\ee

The $\CP$ models possess a $U(1)$ gauge invariance and
have many properties in common with 4D QCD. For example, they are classically scale invariant and have 
classical instanton solutions which, like QCD instantons, are of arbitrary radius. Moreover, the $\CP$ models undergo
dimensional transmutation via a conformal anomaly, acquiring a mass gap and becoming asymptotically free.
The $\CP$ analogy was also used by Witten \cite{Witten79} to support the assertion that, in unbroken, asymptotically
free gauge theories, classical instantons would ``melt'' due to quantum fluctuations and are thus irrelevant to topological
charge structure in QCD. The arguments in both Refs. \cite{Luscher78} and \cite{Witten79}
lead to a picture of the QCD vacuum which is in some respects a four-dimensional generalization of Coleman's original
discussion \cite{Coleman76} of $\theta$-dependence in the massive Schwinger model, in which the $\theta$
parameter appears as a background electric field, and instantons play no role. 
For the comparison of topological charge structure in two-dimensional $U(1)$ theories with that 
in QCD, Luscher argued that a precise 
analogy between the two theories could be made by identifying the Chern-Simons currents,
since in both theories, nonvanishing topological susceptibility implies a $p^2=0$ pole in the $j_{\mu}^{CS}$
correlator. The crucial observation here is that {\it the $U(1)$ gauge potential $A_{\mu}$ in the $\CP$ model
should be identified not with the 4D Yang-Mills gauge potential, but rather with the abelian 3-index Chern-Simons tensor
(\ref{eq:CStensor}).} Similarly, a Wilson loop in 2D corresponds not to a Wilson loop in 4D
but rather, a ``Wilson bag'', i.e. an integral of $A_{\mu\nu\rho}$ over the three-dimensional world volume of a 
membrane-like surface. The surface of this bag separates regions of 
spacetime which have effective values of $\theta$
which differ by $2\pi$ (or by fractions of $2\pi$ for a fractionally charged bag). Here the effective local value
 of $\theta$ is the analog of the local background electric field in Coleman's Schwinger model analysis.
Just as the worldline of a charged particle in 2D serves as a domain wall separating vacua with two different
values of background electric field, so too does the Wilson bag surface in QCD separate two ``k-vacua'' with
values of $\theta$ that differ by integer multiples of $2\pi$. 

\subsection{Theta-Dependence in QCD from string/gauge duality}

  Remarkably, the same physical picture emerges from AdS/CFT duality in the context
of Witten's brane construction of QCD \cite{Witten98}. In this construction, 
the Wilson bag surface associated with the Chern-Simons
tensor corresponds to a wrapped 6-brane in type IIA string theory. (The six brane is wrapped around a compact $S_4$,
so it looks like a 2-brane or membrane in $3+1$ dimensions.) Starting with the string theory on 
$R_4\times S_1\times R_5$, one considers the result of introducing $N$ coincident 4-branes, which are wrapped around
the $S_1$ with supersymmetry breaking boundary conditions. The resulting theory on the branes is thus described (at least
at long distances) by a four-dimensional $SU(N)$ Yang-Mills gauge theory without supersymmetry. The origin of the
$\theta$ term in QCD is a five-dimensional Chern-Simons term on the 4-branes 
of the form $a\wedge F\wedge F$ where a is the $U(1)$
gauge field that couples to the Ramond-Ramond charge of IIA string theory. When the radius of the compactified 
dimension is small, this reduces to a four-dimensional theta term $\theta F\wedge F$ where $\theta$ is given by the Wilson
line of the $RR$ $U(1)$ field around the compact dimension,
\begin{equation}
\label{eq:wl5}
\theta = \oint a_5 dx_5 \,\,.
\end{equation}
In the brane-induced geometry of the IIA string theory, the compact $S_1$ goes around the circumference of a two-dimensional
disk $D$ which has a black hole singularity at it's center. The value of $\theta/2\pi$ given by (\ref{eq:wl5})
determines the number of units of R-R flux that are threaded through the singularity. This picture leads to multiple
``k-vacuum'' states where the local values of $\theta$ differ by integer multiples of $2\pi$.
Adjacent k-vacua are separated by domain walls (to be identified with Wilson bag surfaces) which are the AdS/CFT dual analog of
wrapped 6-branes. The fact that $\theta$ jumps by a multiple of $2\pi$ when crossing a domain wall
(a defining property of the Wilson bag surface integral)
follows in the string theory from the quantization of the R-R charge on the 6-brane.

\begin{figure}
\vspace*{2.0cm}
\includegraphics{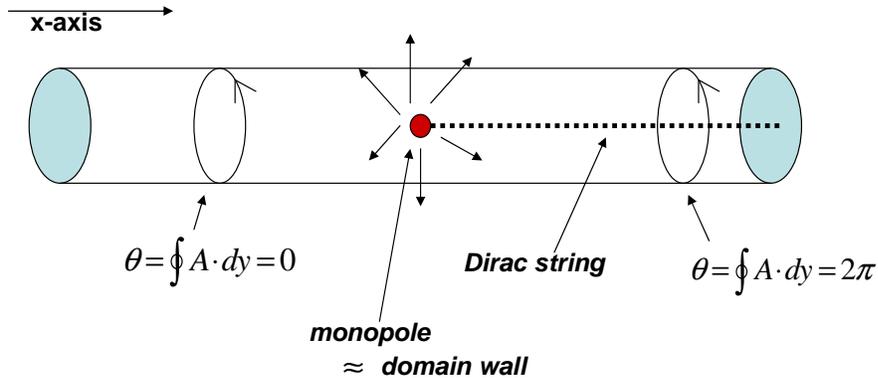}
\vspace{6.5cm}
\caption{Holographic view of a domain wall in (1+1)-dimensional $\CP$ theories from a (3+1)-dimensional perspective.
The long axis of the cylinder becomes the spatial axis of the (1+1)-dimensional theory. Plot is at a fixed time.}
\label{fig:laughlin}
\end{figure}

\subsection{Corbino disks, the integer quantum Hall effect, and thin domain walls}

The interpretation of the 4D theta term as a dimensionally reduced 5D Chern-Simons term is a central feature of the 
AdS/CFT view of theta dependence. The general analogy we are pursuing in this paper suggests that we should interpret
a $\theta$ term in the $CP^{N-1}$ model as a dimensionally reduced three-dimensional Chern-Simons term. In fact, this 
analogy brings out the deep connection between Witten's description of theta-dependence in Yang-Mills theory \cite{Witten98}
and Laughlin's famous topological
interpretation of the integer quantum Hall effect \cite{Laughlin}. The topology of the (2+1)-dimensional superconductor
considered by Laughlin is realized physically by a 
``Corbino disk,'' a 2D disk with a hole in it. The Wilson line integral around the disk measures the magnetic flux 
through the hole. For the purpose of dimensionally reducing this to a (1+1)-dimensional theory with a $\theta$
term, it is convenient to consider the topologically equivalent situation of a long thin cylinder, with units
of magnetic flux going down the center of the cylinder, as depicted in Figure \ref{fig:laughlin}. The quantized flow of Hall current down the length of the 
cylinder corresponds to a change of $\theta$ by an integer multiple of $2\pi$. A domain wall between two different
k-vacua along the cylinder is represented in the higher dimensional theory by a magnetic monopole at that location
with a Dirac string on one side of it. Note that, in this picture, the transition from, e.g. the $\theta=0$ vacuum
to the $\theta=2\pi$ vacuum takes place over a distance proportional to the radius of the cylinder, so that in the dimensionally
reduced theory, the domain wall becomes infinitely thin. In fact, in the brane construction of QCD, the radius of
the compact spatial dimension can be interpreted as a cutoff, somewhat analogous to the lattice spacing in a lattice 
formulation of the 4D theory. This suggests that, if we interpret the coherent sheets observed in the Monte Carlo
simulations as domain wall boundaries between k-vacua, their thickness  
should scale to zero linearly with the lattice spacing, i.e. that they should be roughly the same thickness in lattice 
units, independent of $\beta$. This is just what is observed in QCD configurations \cite{Horvath05_corr}. As we discuss
in Section IV(B), the thickness of the topological charge structures observed in $CP^{N-1}$ also scales to zero linearly in 
the continuum limit.

\subsection{Wilson Lines, Wilson Bags, and Charge Screening}

As emphasized by Witten \cite{Witten79} one should draw a clear distinction
between topological charge structure in a spontaneously broken gauge theory 
(e.g. the 2D $U(1)$ Higgs model) where instantons have a fixed size and are expected to
be relevant in the quantum theory, and the structure of an asymptotically free
theory such as $\CP$, where the multiple k-vacuum/domain wall picture is expected. 
Consider a 2D $U(1)$ gauge theory
in an infinite volume in which $\theta$ is a nonzero constant only on a finite subvolume $V$, surrounded by a region
in which $\theta=0$, i.e. we add a term to the Euclidean action
\begin{equation}
\label{eq:thetaterm}
S\rightarrow S+ \int d^2x \theta(x)\epsilon_{\mu\nu}F^{\mu\nu} \,\,.
 \end{equation}
where $\theta(x)=\theta$ inside $V$ and $\theta(x)=0$ outside.
Upon integration by parts, such a $\theta$ term in the path integral
is equivalent to including a Wilson loop around the boundary of $V$, interpretable as the
worldline of a test charge of $\theta/2\pi$ (in units where the charge of the $\CP$ field is one),
\begin{equation}
\label{eq:wloop}
S\rightarrow S+\frac{\theta}{2\pi}\oint_{C} A\cdot dx
\end{equation}
where $C=\partial V$ \,\,\,      .
 
No matter what the physical mechanism for topological charge fluctuations, we expect the ground state energy to
be periodic under $\theta\rightarrow \theta+2\pi$. However, this periodicity can arise in two physically
distinct ways. In a dilute instanton scenario, the topological charge comes in locally
quantized lumps which are well enough separated from nearby lumps that we can carve out a local
subvolume around a given lump over which the topological charge integrates to an integer $\nu_i$.
(We do not consider models involving highly overlapping instantons.) The local quantization of
topological charge in an instanton model allows periodicity in $\theta$ to be satisfied locally in each small
subvolume, since the $\theta$ term simply multiplies the partition function by periodic factors $e^{i\theta\nu_i}$ for each
instanton. If we change $\theta$ continuously from 0 to $2\pi$, we expect smooth periodic
behavior without any
bulk transition in the vacuum. Expressing the $\theta$ term as a Wilson loop around the boundary, Eq. (\ref{eq:wloop}),
we can assume that, for a dilute instanton gas, as $V$ gets large, the gauge field  
in the asymptotic region can be taken to be pure gauge. In this case,
the precise location of the Wilson loop has no physical significance. As long as it is in the asymptotic pure-gauge
region, it merely counts the number of instantons minus antiinstantons inside $V$. 
A much different situation occurs in QCD-like theories where the gauge invariance is unbroken and a finite mass gap
arises via quantum effects. In these theories
(QCD, discussed in \cite{Horvath03_struct} and $CP^{N-1}$, discussed here), 
Monte Carlo calculations appear to support Witten's arguments 
that instantons disappear from the quantum theory and are, in some sense, replaced by domain walls between k-vacua.
Topological charge does not appear in locally quantized lumps but rather in extended coherent structures of codimension 1.
In this situation, the boundary condition that $F_{\mu\nu}=0$ asymptotically is not the correct one. 
If a Wilson loop around $V$ is included in the path integral, it will introduce a physical domain
wall separating the $\theta=0$ vacuum outside from the nonzero-$\theta$ vacuum inside. Periodicity in $\theta$
arises in a discontinuous way, involving a ``string-breaking'' or charged pair production process, resulting in
the screening or partial screening of the Wilson loop. This is just the mechanism discussed in 
Coleman's original description of $\theta$-dependence in the massive Schwinger
model. 

If the topological susceptibility is nonzero, then for generic values of $\theta$, the free energy per unit 
volume inside $V$, $E(\theta)$ is greater than $E(0)$, the value outside $V$.
The Wilson loop around $V$ thus satisfies an area law,
\begin{equation}
\label{eq:arealaw}
\langle W(C)\rangle \sim \exp\left[-(E(\theta)-E(0))V\right] \,\,.
\end{equation}
This exhibits the linear, confining Coulomb force between test charges of $\pm\theta/2\pi$
at opposite ends of the box. In the two-dimensional case, confinement of U(1) charge and nonvanishing topological susceptibility
both arise from the massless pole
in the Chern-Simons current correlator. In four-dimensional QCD, the analog of (\ref{eq:arealaw}) is not confinement
of quarks, but rather a ``volume law'' for Wilson bags: 
\begin{equation}
 \label{eq:bag}
\left\langle \exp\left[i(\theta/2\pi)\int_S A_{\mu\nu\rho}dx^{\mu}dx^{\nu}dx^{\rho}\right]\right\rangle \sim \exp\left[-(E(\theta)-E(0))V\right]
\end{equation}
where $S$ is the surface of a closed bag and $V$ is the enclosed 4-volume.
In the two-dimensional theory, as $\theta$ is increased the constant electric field
between the test charges increases. As $\theta$ crosses $\pi$, the field becomes strong enough to produce a pair of
charged scalars out of the vacuum and send them to opposite ends of the box, screening one unit of electric flux.
In the $CP^{N-1}$ model the gauge field is actually an auxiliary field composed of $z^+z^-$ pairs, 
so it is more accurate to describe the screening process as resulting from a collective motion of the
charged $z$-particles in the vacuum which leaves a net charge at the two ends of the box.
At $\theta=\pi$, the screened and unscreened vacua are degenerate, with the two vacua containing a background of
$\pm\frac{1}{2}$ a unit of electric flux. As theta goes through $\pi$ there is a sudden transition from the unscreened
to the screened vacuum. [By contrast, a dilute instanton gas leads to a background electric 
field $\propto \sin\theta$, which goes smoothly through zero at $\theta=\pi$.] As $\theta$ is further increased
toward $2\pi$, the energy per unit volume $E(\theta)$ decreases. Finally, at $\theta = 2\pi$ we have $E(2\pi)=E(0)$, and
the area term in the Wilson loop vanishes. The external test charge is completely screened by the polarization of
the vacuum. At this point there is again no net background flux and no force between the test charges. 

A similar discontinuous behavior at $\theta=\pi$ is expected in the case of four-dimensional QCD \cite{Luscher78,Witten79}, where
the Wilson loop that is used to circumscribe the region of nonzero $\theta$ in the two-dimensional theory
is replaced by the Wilson bag, 
which does the same thing in four-dimensional Yang-Mills theory.  
The force between the bag walls vanishes when the bag has integer charge, i.e. when the
step in $\theta$ across the bag wall is $2\pi$ (or an integer multiple of $2\pi$). 
This fully screened bag is the gauge theory version of the wrapped
6-brane of IIA string theory. The vanishing of the force between the walls of the bag for $\theta=2\pi$
is the gauge theory manifestation of the fundamental string theory result, first discovered by Polchinski \cite{Polchinski}, 
relating the quantization of Ramond-Ramond charge on a D-brane and
the vanishing of the force between two D-branes due to closed string exchange.

In order to discuss topological charge structure in this theoretical framework, we 
need to determine what the elementary ``quasiparticle'' excitations of the vacuum
are, and what type of topological charge structure is associated with these excitations.
Note that we are discussing here the excitations in the flavor singlet channel accessed by
the gauge field and topological charge operators, after integrating out the $z$ fields. The 
spectroscopy of the $\CP$ models also includes a multiplet of light nonsinglet mesons (which 
we use here to define the overall mass gap or relative lattice spacing for different $\beta$'s.)
These nonsinglet mesons are the lowest lying physical states in the charge neutral sector.
Unlike the singlet channel, the nonsinglet states can be excited by a local $z_i^*(x)z_j(x)$ operator.
The spectral structure of the flavor-singlet channel can be discussed in terms of the $A_{\mu}$ correlator,
or, equivalently, the Chern-Simons correlator (\ref{eq:cscorr}). [Because of the constraint 
${\bf z}^\dagger \cdot {\bf z} = 1$, the lowest dimension operator that is available to access the flavor
singlet channel is the currrent (\ref{eq:current}), which is proportional to $A_{\mu}$.]
Consider the correlator as an analytic
function of $s= -p^2$. Since the topological charge correlator
$p^2G(p^2)$ is gauge invariant and there is a mass gap in the theory, 
it can be asserted that the imaginary part of $p^2G(p^2)$ is zero below 
some threshold for real particle production. The threshold is the mass of the lightest flavor-singlet
state, which could be either the mass of a flavor-singlet meson, or (if no stable flavor-singlet meson
exists) $s<4M^2$ where $M$ is the mass of the lightest nonsinglet meson. However, because of the zero mass
pole in the Chern-Simons correlator, the imaginary part of $G(p^2)$ includes a zero mass delta function
$\propto \delta(s)$. 
This analytic structure in the gauge correlator for $\CP$ models has a close analog in superconductivity theory \cite{Tinkham}.
In that case, the complex conductivity for any nonzero frequency $\omega$ below the mass gap has no absorptive (real)
part, due to the absence of resistance. However, because of the purely accelerative mode of the Cooper pairs,
the dispersive part has a $1/\omega$ pole, 
and the absorptive part has a $\delta(\omega)$, representing the DC flow of supercurrent.
Because of the presence of the supercurrent, if a charged electron is inserted 
into a superconductor, the quasiparticle that forms is
electrically neutral, corresponding to the electron plus the backflow of the Cooper pairs, which
screen the electron charge \cite{Kivelson}. The excess charge from the electron ends up on the surface of the superconductor.
Thus a quasiparticle excitation in a superconductor is an electrically neutral screened electron. 
Following this analogy combined with the preceeding discussion, we propose that the coherent topological 
charge excitations observed in $CP^{N-1}$ and QCD should be identified with the screened Wilson loop 
and the screened Wilson bag, 
respectively. In the $CP^{N-1}$ case, the screened Wilson line can be interpreted as the world line of a charged $z$-particle whose
Coulomb field has been cancelled out by the backflow of charge in the vacuum. Thus the gauge field
associated with an elementary excitation is a one-dimensional thread of $A_{\mu}$ flux which is constant along its length and
whose transverse cross-section is a delta-function. On the lattice, this corresponds to the coherent excitation
of a single line of links. Recall that the $A_{\mu}$ field is an auxiliary field whose equation of motion sets
it equal to the $z$-particle current, Eq. (\ref{eq:London}).
Thus a Wilson loop excitation can also be interpreted as a filamentary current flow. 

\begin{figure}
\begin{center}
\epsfxsize=.60\textwidth
\epsfbox{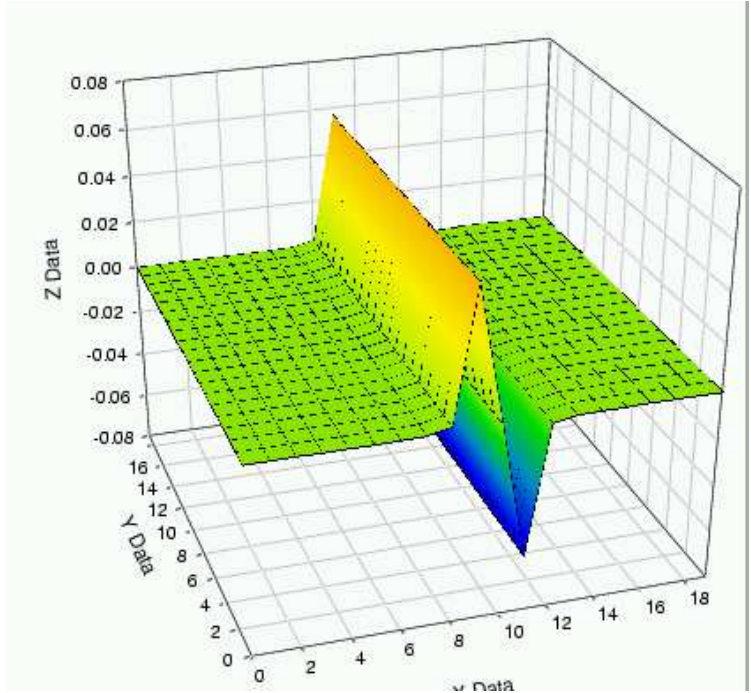}
\end{center}
\caption{Plot of the overlap topological charge distribution for a single Wilson line excitation.}
\label{fig:wline}
\end{figure}

In Figure \ref{fig:wline} we show the topological charge distribution
calculated by the overlap method for a gauge field which consists of a single straight Wilson line excitation. The 
links along the Wilson line are taken to be a constant nonzero phase, with all other links on the lattice set to unity. 
The topological charge distribution associated with the Wilson line excitation is a dipole layer, with
oppositely charged one-dimensional coherent regions on either side of the Wilson line. This resembles 
the local structure observed in the Monte Carlo distributions. 
In a similar way, a Wilson bag excitation of the Chern-Simons tensor which is constant on a 3-surface (and zero
elsewhere) in four-dimensional QCD yields
a topological charge distribution consisting of oppositely charged coherent three-dimensional regions on either
side of the bag surface. Again, this provides a reasonable description of the coherent structure seen in 
the QCD Monte Carlo distributions.

\section {Lattice $CP^{N-1}$ Models and Monte Carlo Calculations} 
\label{sec:lattice}

\subsection{Lattice Action}

For the lattice formulation of the $CP^{N-1}$ models, we introduce $U(1)$ link fields $U(x,x+\hat{\mu})$ in the usual way and take the action to 
consist of gauge  invariant nearest-neighbor hopping terms,
\be
\label{eq:action}
S = -\beta N\sum_{x,\hat{\mu}}{\bf z}(x)^\dagger \cdot {\bf z}(x+\hat{\mu})U(x,x+\hat{\mu})\,+\,c.c. \,\,.
\ee

The naive ultralocal definition of  
topological charge density on the lattice is in terms of the plaquette phase:
\be
\label{eq:logplaq}
q_P(x) = \frac{1}{2\pi i}\log U_P(x) 
\ee
where $U_P(x)$ is the product of phases around a plaquette at site $x$. Here the principle branch of the log is chosen
so that the charge on a plaquette always lies between $-\frac{1}{2}$ and $+\frac{1}{2}$. With toroidal boundary
conditions, this definition sums to an integer-valued global topological charge.
The analogous construction of topological charge in four-dimensional QCD from gauge variables on the lattice has been given by Luscher
\cite{Luscher82}.  

We have found that, just as in four-dimensional QCD, a much better definition of the local topological charge density is given
in terms of an exactly chiral overlap Dirac operator D \cite{Neuberger}. As shown in \cite{Hasenfratz98}, if $D$ satisfies
Ginsparg-Wilson relations, then the lattice topological charge density operator which appears in the axial U(1)
anomaly equation is 
\be
\label{eq:overlapq}
q(x) = \frac{1}{2} tr\gamma_5D(x,x)
\ee
where the trace is over spin  indices in $\CP$ and over spin and color indices in QCD. 
Using the construction of the overlap operator D described in
the next section, we have studied topological charge distributions in $CP^1$, $CP^3$, and $CP^9$.
We find that the density $q(x)$ defined in (\ref{eq:overlapq}) reveals the coherent long-range structure
that was obscured by the short range noise inherent in the ultralocal operator $q_P(x)$. A detailed comparison
of distributions obtained using the overlap $q$ with those obtained using the plaquette operator $q_P$
has been carried out and will be presented elsewhere.

\subsection{Monte Carlo Calculations} 

The lattice $CP^{N-1}$ action, Eq. (\ref{eq:action}) was used for Monte Carlo 
simulation. The updating of the ${\bf z}$ fields was done by a Cabibbo-Marinari
algorithm consisting of a sequence of $SU(2)$ heat bath updates applied to
all possible pairs of $z$-components. The gauge links were updated by
a multi-hit Metropolis algorithm. Calculations were done for $CP^1, CP^3$, and
$CP^9$. For each value of $\beta$, we determine a value for the mass gap by
studying the exponential falloff of the $z_i^*z_j$ meson correlator for $i\neq j$,
\be
\int dx_2\langle z_i^*z_j(x)\,z_j^*z_i(0)\rangle \,\sim\, const.\times e^{-\mu x_1} \,\,.
\ee
The evaluation of $\mu$ determines the mass scale. In Table \ref{tab:massgap} we give the mass gap
$\mu$ in lattice units for various values of $N$ and $\beta$. 
The Monte Carlo routine was checked extensively by comparing results with
the strong coupling expansion \cite{Seiberg} to order $\beta^8$ for both the meson correlator and also for the topological
charge correlator for the ultralocal $q_P(x)$ operator.

\begin{table}[h]
\centering
\caption{Massgap in lattice units}
\begin{tabular}{||c|c||c|c||c|c||} \hline \hline
$\beta$      & $CP^1$    & $\beta$      & $CP^3$           & $\beta$    & $CP^9$  \\ \hline \hline
1.0	     & .438(5)   & 0.8          & .554(2)          & 0.7	& .406(2) \\ \hline
1.1          & .286(5)   & 0.9          & .327(3)          & 0.8        & .212(2)    \\ \hline
1.2          & .179(3)   & 1.0          & .180(1)          & 0.9        & .0895(6)     \\ \hline
1.3	     & .111(1)	 & 1.1	        & .0882(7)         & 1.0	& .0579(2)   \\ \hline
1.4	     & .0696(8)  & 1.2          & .0531(3)         & 1.1        & .0475(4)   \\ \hline \hline
\end{tabular}
\label{tab:massgap}
\end{table}

We study lattice volumes up to $50^2$. For each $N$, values of $\beta$ were chosen to
cover a range of correlation lengths from approximately $\xi=\mu^{-1}=3$ to 20.
Correlator fits were carried out by standard methods, using covariant $\chi^2$ 
minimization. The statistical errors and autocorrelation
times were determined by a bootstrap algorithm.

\subsection{Overlap Dirac Operator} 
\label{sec:overlap}

The overlap construction \cite{Neuberger} provides a prescription for constructing an
exactly chiral Dirac operator $D$ satisfying the GW relations (\ref{eq:GW}). The construction
begins with a suitable ultralocal discretization of the Dirac operator as a kernel. Here we will
use the usual Wilson-Dirac operator as the kernel, 
\be
D_W = \frac{1}{2} \gamma_\mu (\nabla_\mu + \nabla_\mu^*)- 
	\frac{1}{2} a \nabla^*_\mu \nabla_\mu \,\, ,
\ee
where $\nabla_\mu$ and $\nabla_\mu^*$ are the forward and backward 
lattice derivatives, respectively, 
\bea
\nabla_\mu \psi(x) &=& \frac{1}{a} \left( U_\mu(x) \psi(x+a{\hat \mu}) 
	- \psi(x) \right) \\
\nabla_\mu^* \psi(x) &=& \frac{1}{a} \left( \psi(x) 
	- U^{\dagger}_\mu(x-a{\hat \mu}) \psi(x-a{\hat \mu}) \right) \,\, .
\eea
The overlap operator can be written as 
\be
D = \frac{1}{a} \left( 1 + \gamma_5 \, \epsilon( H_W(m) ) \right) \,\, ,
\ee
where $H_W(m)=\gamma_5 D_W(-m)$ and the sign function is 
\be
\epsilon( H_W(m) ) = \frac{H_W(m)}{\sqrt{H^{\dagger}_W(m) H_W(m)}} \,\, .
\ee
This operator has a generalized chiral symmtry given by the 
Ginsparg--Wilson relation
\be
\gamma_5 \, D + D \gamma_5 = {\bar a} D \, \gamma_5 D \,\,,
\ee
as is easily verified.

The Wilson mass parameter can 
be chosen to lie in the range $0<m<2$ with the various values of 
$m$ giving the same continuum physics.  This range is  
allowed, at least in the case of free fields and for sufficiently smooth gauge 
field configurations. We have carried out our calcuations in the range 
$0<m<1$ and found the results to be insensitive to the choice of $m$ in
this range.
For the lattice sizes used in this study (up to $50\times 50$), it was possible to construct the
overlap operator exactly, using a LAPACK singular value decomposition routine.
(See \cite{Rebbi} for a discussion.)

\section {Topological Charge Structure in $\CP$ models} 
\label{sec:Topcharge}

\begin{figure}
\vspace*{1.0cm}
\includegraphics{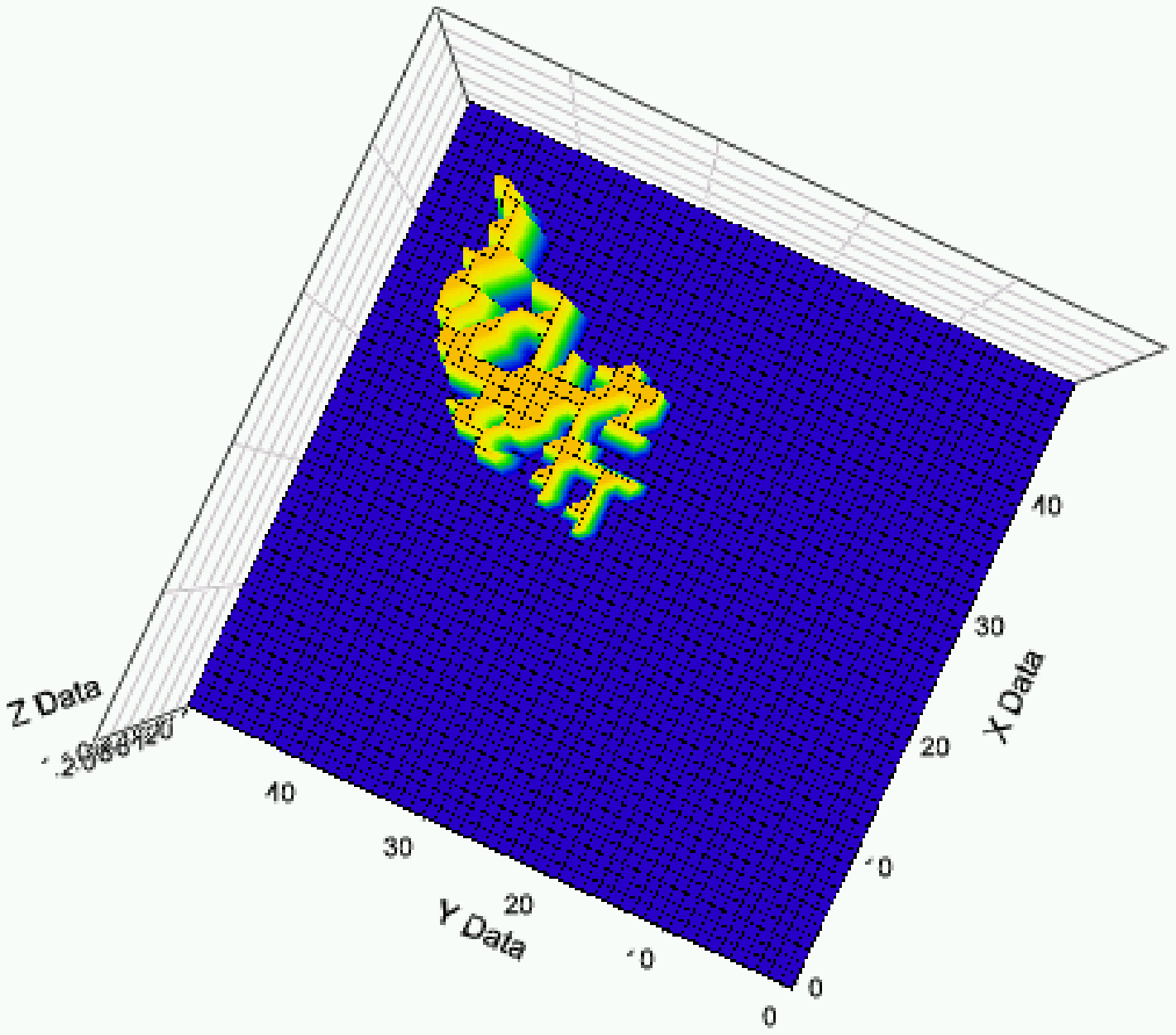}
\includegraphics{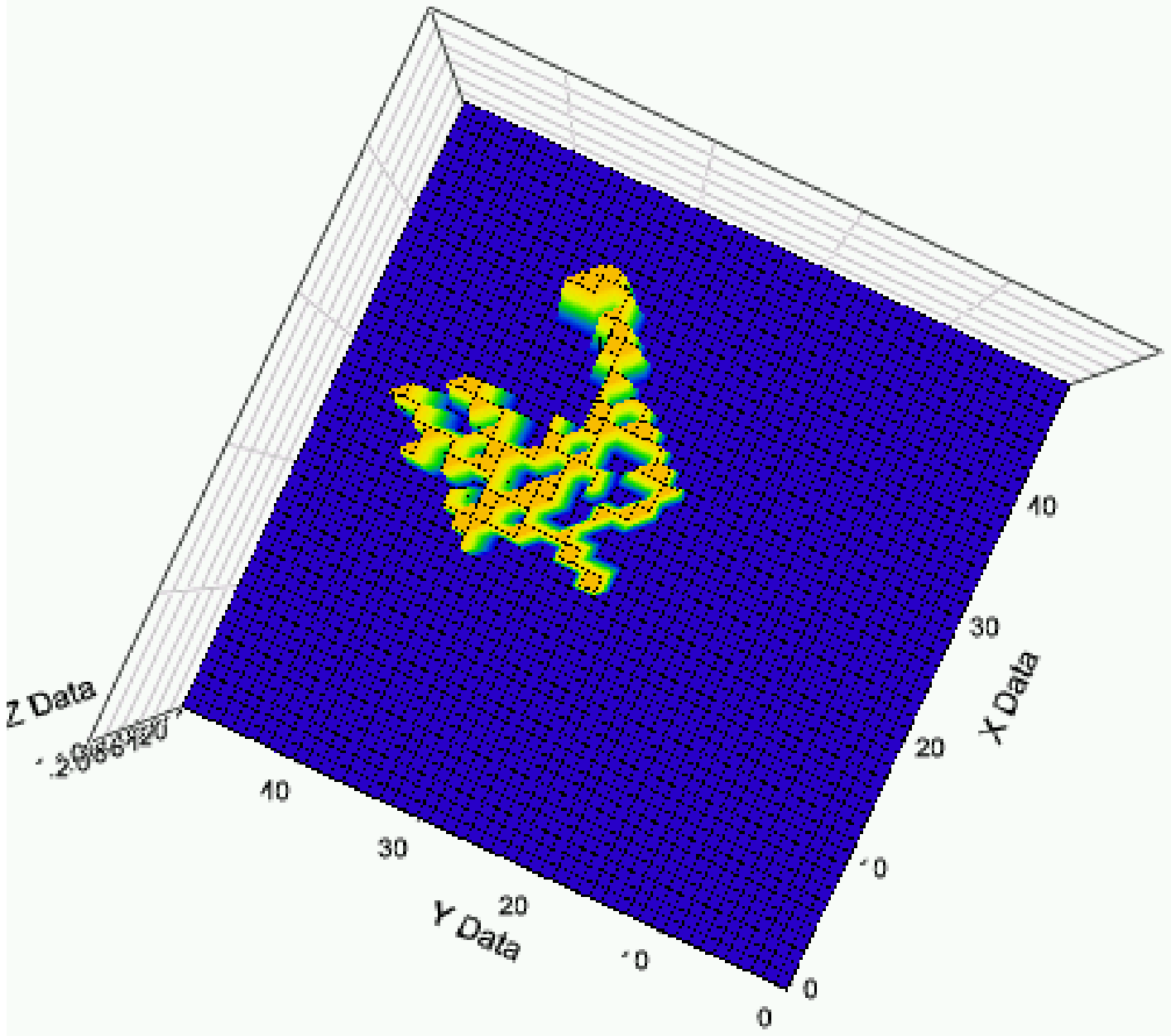}
\vspace{6.5cm}
\caption[]{Two typical largest structures for random $q(x)$ distributions on a $50\times 50$ lattice.}
\label{fig:randomstructures}
\end{figure}

\begin{figure}
\vspace*{1.0cm}
\includegraphics{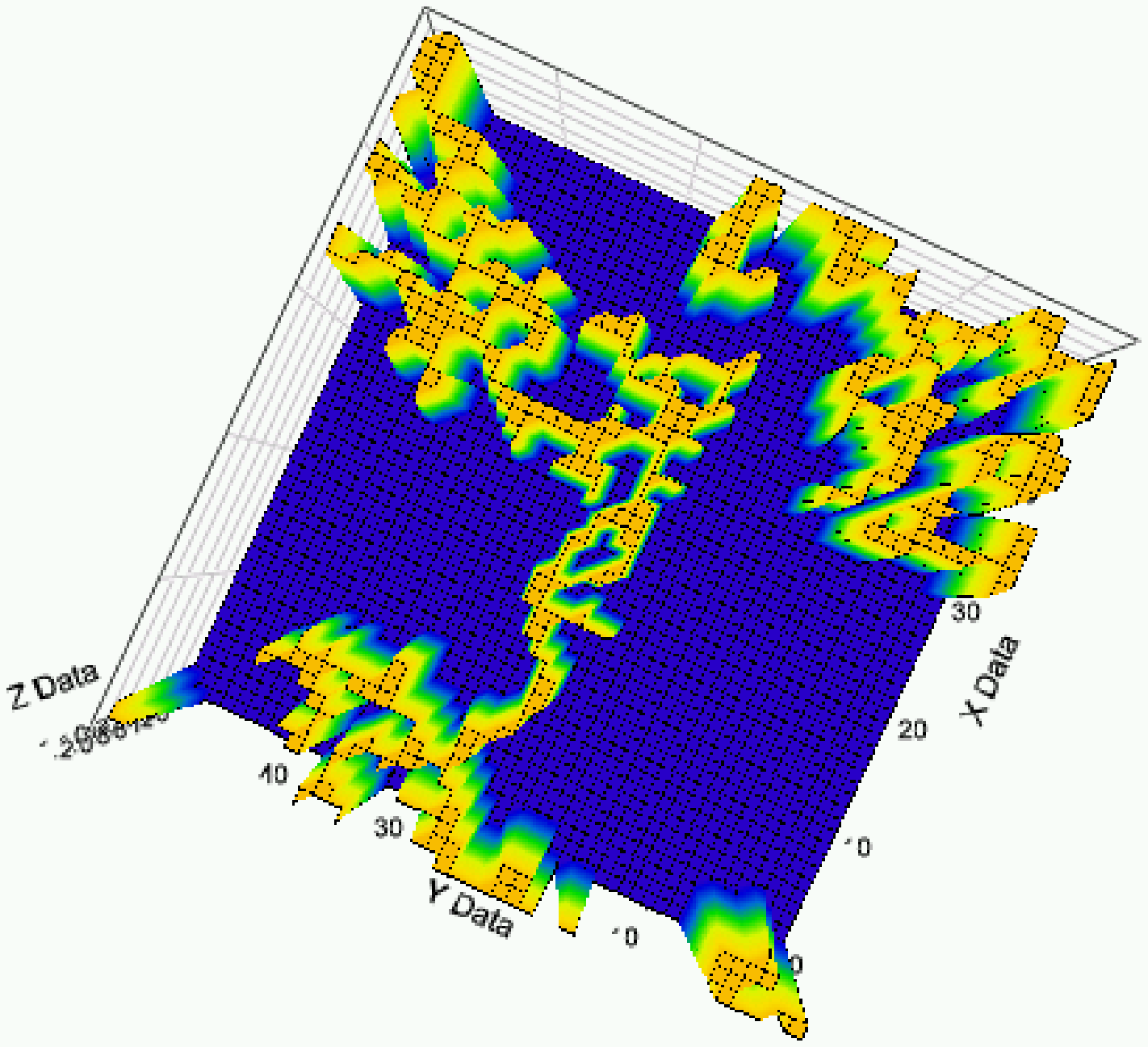}
\includegraphics{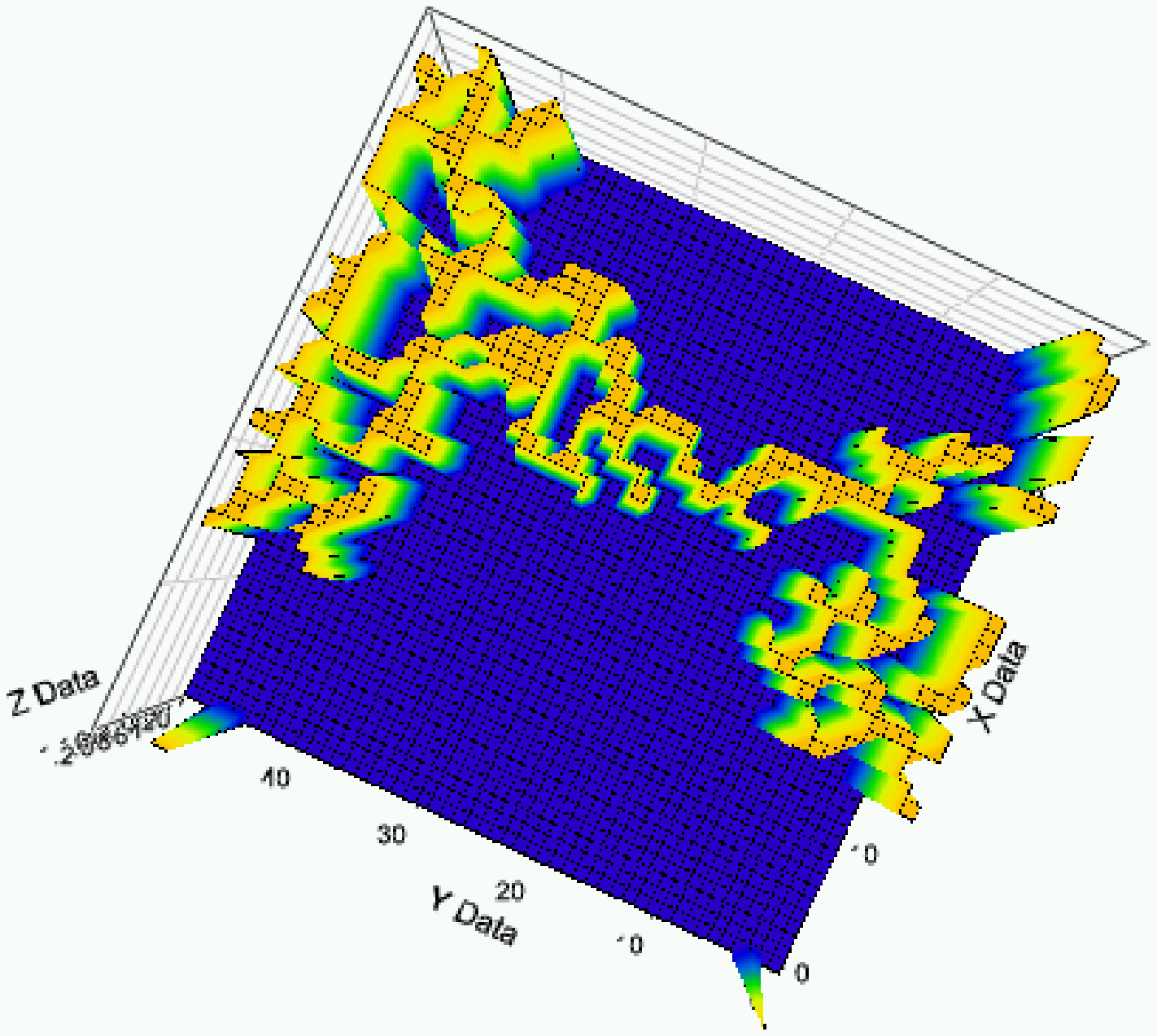}
\vspace{6.5cm}
\caption[]{Two typical largest structures for $CP^3$ on a $50\times 50$ lattice at $\beta=1.2$.}
\label{fig:CP3structures}
\end{figure}

\begin{figure}
\vspace{-1.0cm}
\begin{center}
\epsfxsize=0.80\textwidth
\epsfbox{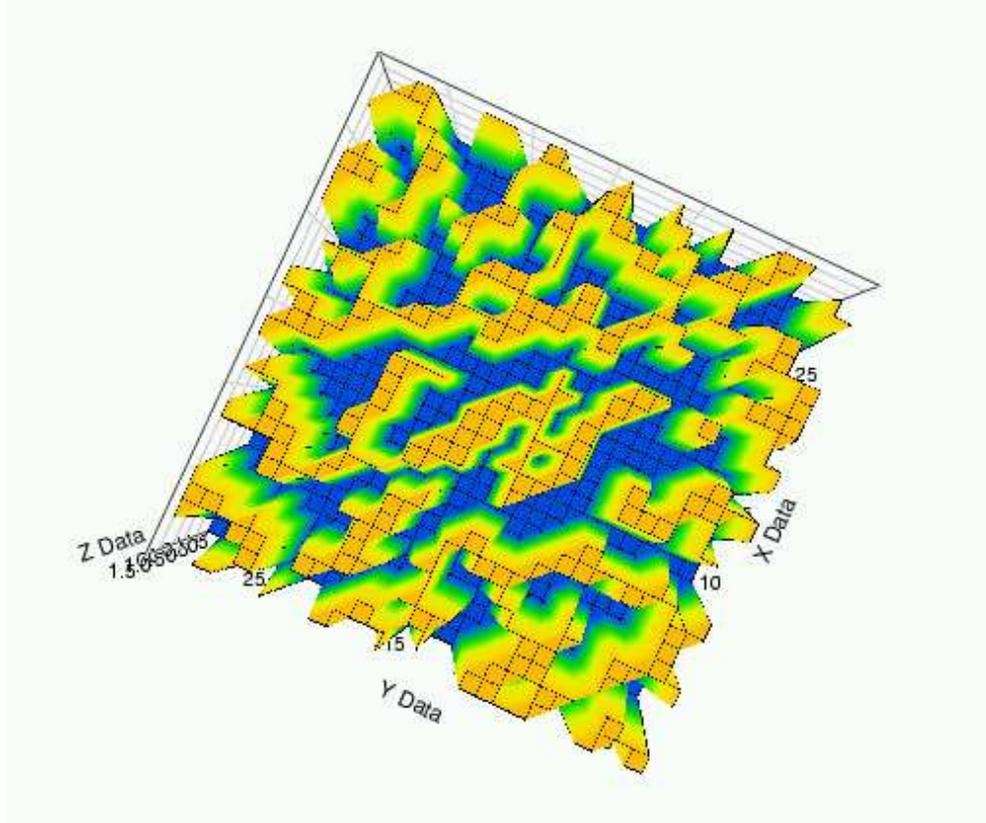}
\end{center}
\vspace{-0.5cm}
\caption{Plot of the function $sign(q(x))$ for a $CP^3$ configuration on a $30\times 30$ lattice at $\beta=1.2$.}
\label{fig:labyrinth}
\end{figure}

Using the overlap definition of the topological charge, visual inspection of the TC distributions
for individual configurations reveals well-defined ``stringy'' patterns in the form of locally 
one-dimensional long range sign coherent structures in the two-dimensional space. These structures are completely analogous to the
3D sheets found in four-dimensional QCD, and precisely what is expected from the Chern-Simons tensor analogy. 
It is worth pointing out that, when we use the log-plaquette 
definition of topological charge, we do not observe any such sign coherent structures.
[It is interesting to note, however, that much of the qualitative structure seen in the overlap $q(x)$
distribution can also be discerned from the $q_P(x)$ distribution after a simple smoothing
procedure, defining the charge at a site to be the average of $q_P$ for the four plaquettes
around the site. This comparison will be discussed in detail elsewhere.] 

A direct way to get a qualitative view of the coherent topological charge structure in
$\CP$ is to plot the largest connected structure in each configuration. Here we define a connected structure
as a set of nearest-neighbor-contiguous lattice points with the same sign of topological charge.
As a reference, we first study the structure plots for randomly generated distributions. 
Two typical largest structures from a random TC distribution on a $50\times 50$ lattice  
are shown in Fig. \ref{fig:randomstructures}. These
are to be compared with the structures shown in Fig. \ref{fig:CP3structures} 
which are obtained from the overlap TC distribution
for typical $CP^3$ Monte Carlo configurations at $\beta=1.2$ (correlation length $\approx 19$). Overall, the $CP^3$ structures
are much larger in extent (typically as large as the lattice itself) compared with the random structures.
Even more striking is the ``stringiness'' of the $CP^3$ structures. They are characterized by long slender
regions of coherence which are locally one-dimensional. This contrasts with the random structures which
are not only smaller in extent but much more two-dimensional. 

Another qualitative feature that can be illustrated graphically is the layered nature of the topological 
charge distributions, with alternating sign coherent regions interleaved in a somewhat labyrinthine arrangement.
Figure \ref{fig:labyrinth} shows a plot of $f(x)=sign(q(x))$ on a $CP^3$ configuration on a $30\times 30$
lattice. As is the case in QCD, the presence of
thin alternating-sign coherent regions of codimension one is in some sense the maximum amount of long range order
allowable by the required (and observed) negativity of the correlator for nonzero separation.    

To construct a quantitative measure of coherence, we determine the
inverse participation ratio (IPR) defined as the inverse fraction of the
lattice volume occupied by coherent structures, i.e.
$IPR(n)=V/V(n)$, where $V$ is the total volume and $V(n)$ is the
volume occupied by $n$ largest coherent structures. (Thus, small localized structures
give a large IPR, while a structure occupying most of the lattice would give an IPR
close to 1.)

\begin{figure}
\centering
\includegraphics[width=4in,angle=270]{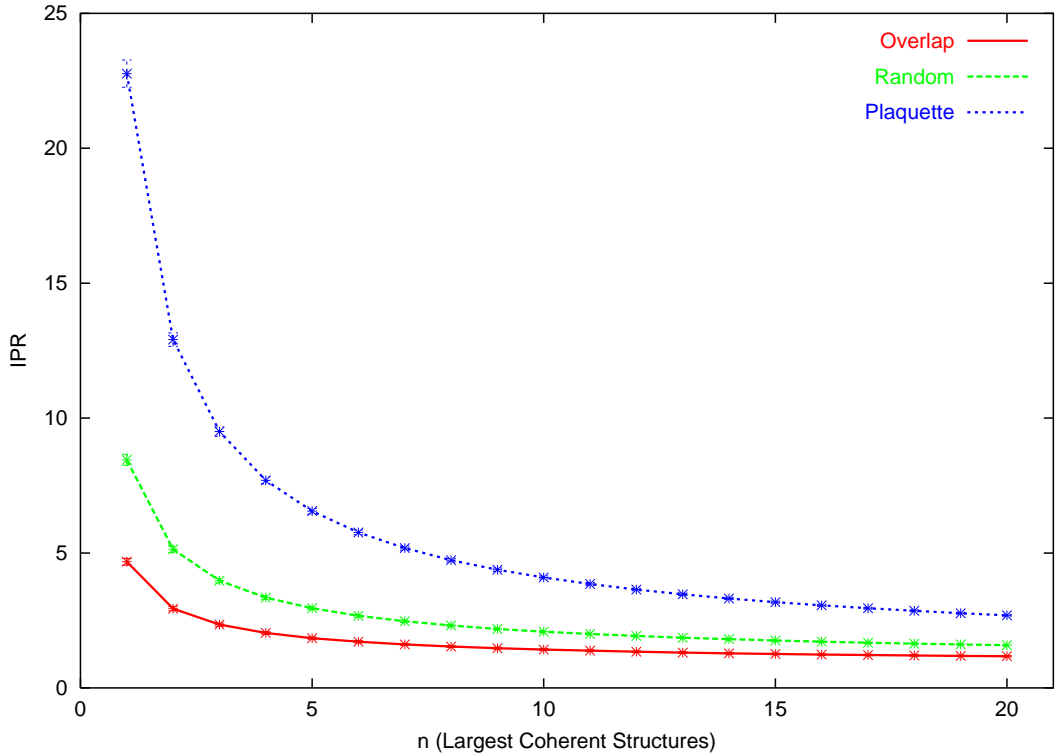}
\caption{Inverse participation ratio (IPR) for the overlap and plaquette
distributions of the topological charge. For a comparison, we show the
result for a random distribution of numbers.}
\label{fig:VolComp}
\end{figure}

Fig. \ref{fig:VolComp} shows the results for both the overlap $q(x)$ distribution and the log-plaquette operator $q_P(x)$. 
 Also shown for comparison is the same
plot for a set of random configurations. 
These results are from a large ensemble of $CP^3$
configurations on a $40\times 40$ lattice with $\beta=1.0$
(correlation length$\approx 5$).
We see that the overlap
definition of $q(x)$ exhibits a clear indication of coherence, e.g.
the typical largest structures are much larger than those in a random
configuration. Somewhat surprisingly, the plaquette phase definition actually
exhibits {\it less} structure than the purely random distributions. This is an effect
of the nearest-neighbor anticorrelation for the plaquette phase.

\subsection {Topological Charge Correlator} 

\begin{figure}
\centering
\includegraphics[width=4in,angle=270]{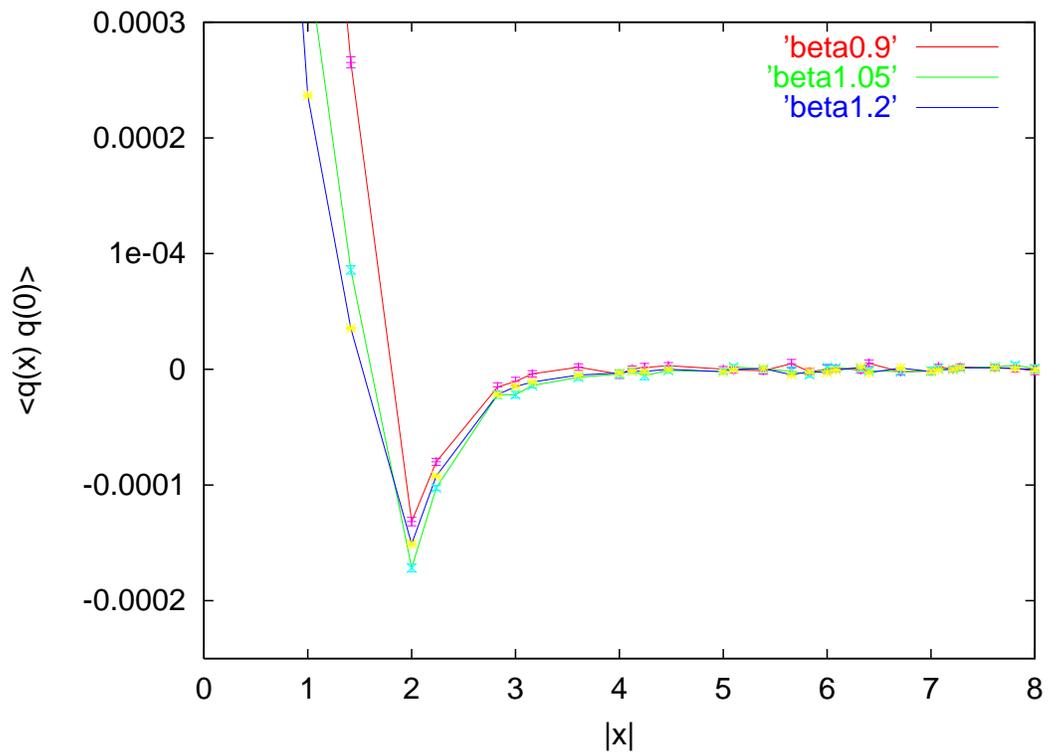}
\caption{Topological charge correlator for $CP^3$ (lattice units).}
\label{fig:tcc_lat}
\end{figure}

In the continuum, the Euclidean topological charge correlator must be negative outside of a
positive contact term at $x=0$. On the lattice, the overlap $q(x)$ is not ultralocal, but it can
be argued that it becomes local in the continuum limit, at least for sufficiently smooth gauge
fields \cite{Luscher78}. Spectral arguments only require the correlator to be negative
when the two operators are non-overlapping. The correlator $\langle q(x)q(0)\rangle$ is shown
in Figure (\ref{fig:tcc_lat}) for $CP^3$ for several values of $\beta$. 
We see that the correlator consists of a 
positive core at $x\leq \sqrt{2}$, and a negative short-range tail starting at $x=2$. 
Note that the correlator is plotted
in lattice units not in physical units. Thus, for example, the location of the minimum of the correlator
is at $x =$ 2 lattice spacings, independent of $\beta$. 
Also, in physical units, the y axis of the plot
would be rescaled by a factor of $1/\mu^4$, so the minimum at $x=2$ is in fact getting much deeper at large
$\beta$, indicating the development of a short-distance power law singularity.

\begin{figure}
\centering
\includegraphics[width=4in,angle=270]{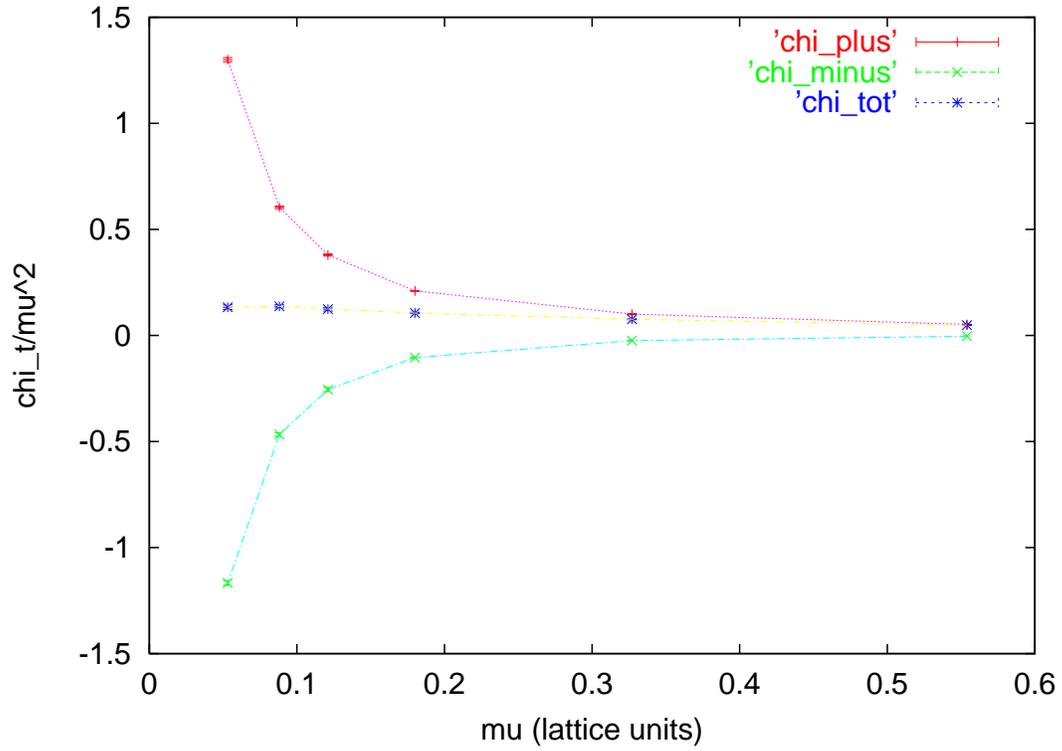}
\caption{Scaling behavior of the positive and negative contributions to the topological susceptibility 
for $CP^3$}
\label{fig:chi_pn}
\end{figure}

In Figure (\ref{fig:chi_pn}) we show the separate scaling behavior of the positive and negative contributions to the topological
susceptibility, i.e. the integral over $\langle q(x)q(0)\rangle$ for $|x|<2$ and for $|x|\geq 2$.
This shows that the contribution of the contact term and the contribution of the negative tail
are separately divergent in the continuum limit, but that the divergence cancels and the topological
susceptibility scales nicely. The spatial extent of the positive core region
is clearly related to the thickness of the coherent regions, while the negative short-range piece
arises from the layered, alternating-sign structure of the configurations. 

\begin{figure}
\centering
\includegraphics[width=4in,angle=270]{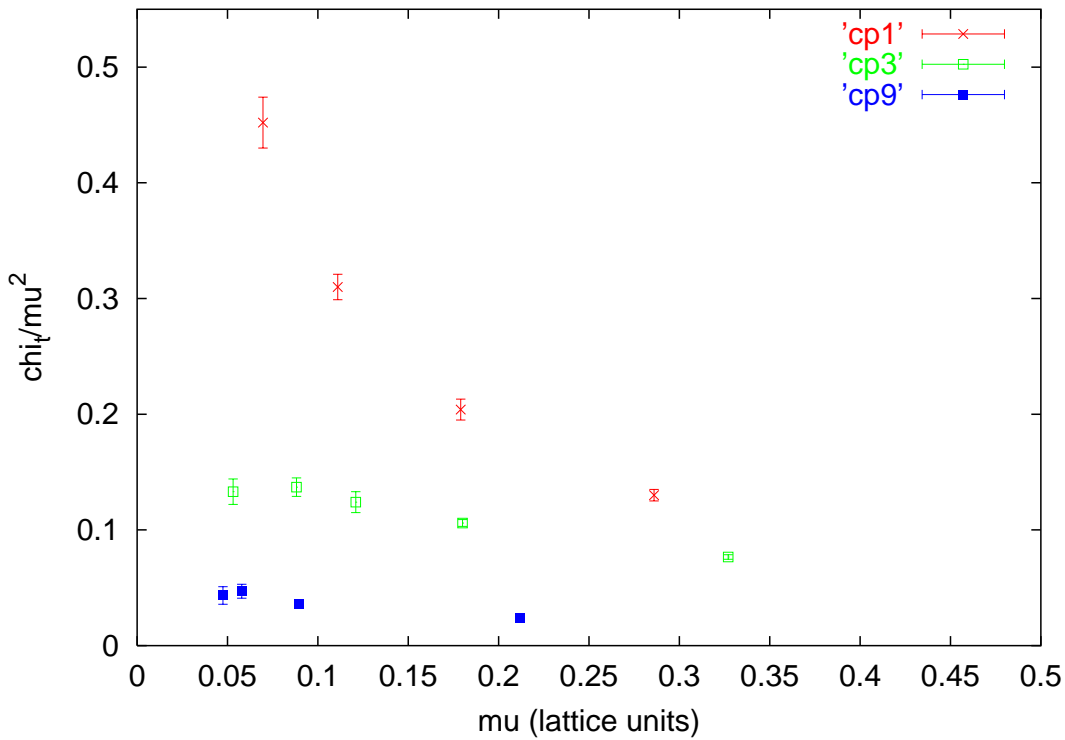}
\caption{Scaling behavior of $\chi_t/\mu^2$ as a function of inverse correlation length
(= mass gap in lattice units) for $CP^1$, $CP^3$ and $CP^9$}
\label{fig:chi_t_scaling}
\end{figure}

Figure (\ref{fig:chi_t_scaling}) shows the full topological susceptibility as a function of the mass gap $\mu$
for $CP^1, CP^3$, and $CP^9$. Here $\chi_t$ is plotted in physical units (dividing the lattice value
by $\mu^2$), so that a constant value indicates proper scaling behavior. 
We observe that $\chi_t$ appears to be properly scaling with the mass gap
for both $CP^3$ and $CP^9$. On the other hand, for $CP^1$ the topological susceptibility is not
even approximately scaling. The anomalous scaling behavior of $\chi_t$ for $CP^1$ is believed to be a consequence
of the divergent contribution of small instantons with radius of order $a$ \cite{Luscher_CP1}. Because of this odd scaling
behavior for $CP^1$, we have focused most of our structure studies on $CP^3$ and $CP^9$. 

\subsection{Thickness of structures}

To support the assertion that the size of the positive contact term in the correlator is determined by the 
thickness of the coherent regions, we can compare this size with a direct measure of the thickness.
We calculate the average thickness of a given coherent structure as follows:(1) Choose a particular point 
on the structure; (2) Walking along a straight path in each of the four directions from that point, measure
the length $l_{\min}$ of the shortest path out of the coherent region, i.e. the path to the nearest opposite-sign
point; (3) Average over all points on the structure. This length of the shortest path should average
to 1/2 the thickness of the structure, so we define the thickness to be $t=2\langle l_{min}\rangle$.
Now let us define $x=x_c$ to be the crossover point where the correlator turns from positive to negative.
In practice, we have estimated this value by linearly interpolating between the positive value of the
correlator at $x=\sqrt{2}$ and the negative value at $x=2$.
If the positive core of the correlator arises from the coherent structures, we might expect that the 
thickness of the structures would be roughly $2x_c$.

\begin{figure}
\centering
\includegraphics[width=4in,angle=270]{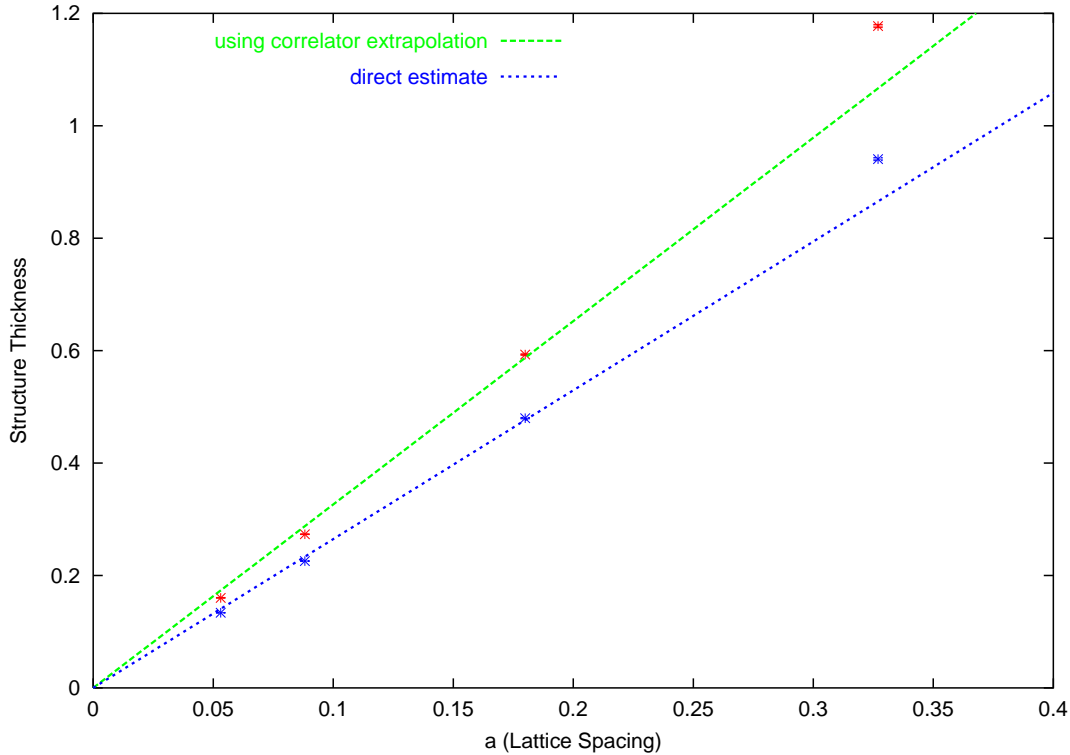}
\caption{Physical thickness of the structures as a function of lattice 
spacing.}
\label{fig:thickness}
\end{figure}

As shown in Figure (\ref{fig:thickness}) the direct estimate of the thickness
$t$, and the value obtained from the core size of the correlator are in approximate agreement. Moreover,
both of these estimates give a thickness which is approximately constant in lattice units and thus scales
to zero linearly in physical units. This agreement leaves little doubt that the positive contact term
in the TC correlator arises from the presence of extended, one-dimensionally coherent topological charge structures 
whose thickness scales to zero in the continuum limit. 

\subsection{Hausdorff dimension of structures}

To construct another quantitative measure of the effective dimensionality of the largest sign-coherent 
structures, we computed their Hausdorff dimension. Starting with each 
site on a structure, we measured the number of other sites $N(r)$ 
on that structure within a radius $r$.  By fitting $N(r)\propto r^d$, 
we extract the Hausdorff dimension, $d$. 
 
Computing the topological charge using the overlap operator, and measuring
the Hausdorff dimension of the largest structure in each configuration
in the ensemble, we obtain
$d=1.26(6)$, confirming the visual impression that these structures are
approximately one-dimensional. For comparison, we studied spin domains
in the two-dimensional Ising model just above $T_c$, adjusted to give structures of
the same volume as the $\CP$ configurations. The Hausdorff dimension
of the Ising spin domains is found to be $d=1.86(5)$.

\subsection{Inherently Global Nature of the Structures}

\begin{figure}
\centering
\includegraphics[width=4in,angle=270]{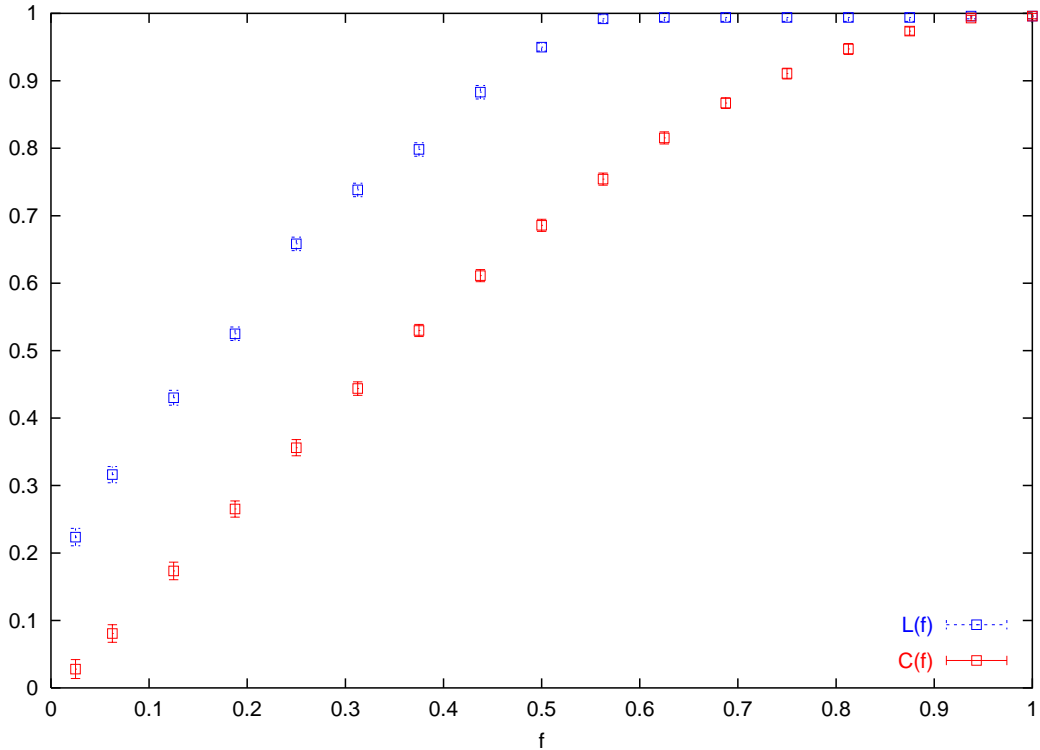}
\caption{Variation of $C(f)$ and $L(f)$ as a function of $f$}
\label{fig:length}
\end{figure}

One result of the QCD studies \cite{Horvath05_global} was that the 3D coherent sheets of 
topological charge are {\it inherently global} in structure, in the sense that the
topological charge is distributed more or less uniformly throughout the structure
and not in localized lumps.
A close observation of the $\CP$ topological charge distributions reveals that the structures in this case are
also inherently global. The 1D structures are, in fact, composed of continuous chains of 
mountains or valleys of almost constant heights. In other words, the most 
intense points join together to form the structures.
To prove this point quantitatively, we study the variation of the length of the largest 
structure and the topological susceptibility as a 
function of the fraction $f$ of points included, starting with the most intense points, as ranked by 
$|q(x)|$. This type of analysis was applied to the QCD structures in Ref. \cite{Horvath05_global,Horvath05_formalism}.
The cumulative function $C(f)$ of the topological susceptibility \cite{Horvath05_global}
represents the fraction of the total topological susceptibility obtained 
when only a fraction $f$ of the most intense points are included in the 
calculation. It is defined as:
\begin{equation}
C(f) = \frac {\chi (f)} {\chi (1)}; \quad \chi (f) \equiv \frac {\left< Q^2 (f)\right> -
\left< Q(f) \right>^2} {V};  \quad
Q(f) = \sum_{x \epsilon S(f)} q(x)
\end{equation}
where $V \equiv a^2 N$, $N$ being the total number of points on the lattice,
and $S(f)$ is the set of points above the threshold introduced via $f$.
The length of a structure is defined as the maximal distance between two
points on the structure, i.e. $l(\Gamma) \equiv \{ max \left| x-y \right|: x,y 
\epsilon \Gamma \}$, where $\Gamma$ is the set of points lying on the same structure.

The ratio of the length of the largest structure at any fraction $f$ to 
that at $f=1$ is $L(f)$, i.e. $L(f) \equiv \frac {l(f)} {l(1)}$.
A plot of $C(f)$ and $L(f)$ (see Fig. \ref{fig:length}) as a function of $f$ 
shows that $L(f)$ increases
much more rapidly than $C(f)$, and reaches $1$ at $f=0.5$,
when only half of the total points are considered on the basis of the
intensity criterion. This proves that the most intense points are connected together 
and the structures are in fact inherently global, in the sense discussed in \cite{Horvath05_global}.

\section{Conclusions and Discussion}

In this paper we have presented Monte Carlo results for topological charge structure in
two-dimensional $\CP$ sigma models. These models exhibit long-range
one-dimensionally coherent topological charge structure which is precisely analogous to 
the three-dimensional coherent sheets observed in four-dimensional pure glue QCD in Reference \cite{Horvath03_struct}.
The analogy between the two-dimensional U(1) gauge potential $A_{\mu}$ in $\CP$ and the {\it abelian} 3-index
Chern-Simons tensor $A_{\mu\nu\sigma}$ in QCD provides a natural framework for interpreting the 
long range structure in both theories. In this framework, the elementary topological charge excitations
of QCD are the ``Wilson bags'' first suggested by Luscher \cite{Luscher78}. A closed 
Wilson bag in four dimensions is analogous in $\CP$ to the creation and annihilation of a $z^+z^-$ pair, forming a 
closed Wilson loop. 
 
The alternating-sign layered arrangement of the coherent regions in the Monte Carlo configurations 
(c.f. Fig. (\ref{fig:labyrinth})) is a central feature of the topological charge structure in both
two-dimensional and four-dimensional theories. At a calculational level, it is clear that this layered structure is enforced by
the requirement that, in the continuum, the two-point TC correlator is negative for nonzero separation.
An attractive feature of the Wilson bag (Wilson line) as the fundamental topological excitation of QCD ($\CP$)
is that the associated topological charge distribution is a dipole layer, which leads naturally to 
the alternating-sign layering that is observed. The physical vacuum is presumably a condensate of 
these surfaces. Both the thickness of the surfaces and the average spacing between adjacent surfaces
are going to zero in the continuum limit, leading to the distinctive features of the TC correlator:
(1) A positive, divergent contact term, (2) A negative, divergent short distance term, and (3) A cancellation
between the (separately divergent) positive and negative contributions to the integrated correlator, giving
a finite topological susceptibility which scales properly, $\propto \mu^2$. For the $\CP$ case, 
we can also associate this structure with the
fact that the gauge field is an auxiliary field representing a coherent oscillation of charged $z$-particles.
The dynamically generated kinetic term for the gauge field, which arises from closed $z$-loops, is in fact
responsible for the $1/q^2$ pole in the CS correlator and hence is the origin of finite topological susceptiblity.

Thus, the picture of the gauge field vacuum as consisting of a dense gas of Wilson line excitations in two-dimensional 
Euclidean space is apparently compatible with the more familiar view of the $\CP$ ground state 
as a plasma of flavor-singlet $z^+z^-$ pairs which supports the oscillations of the
gauge field, generates the $1/q^2$ pole via loop effects, and produces finite topological susceptibility. 
This latter view is made explicit in the large N solution. Although the model of topological charge excitations
based on screened Wilson lines seems to be generally compatible with the large N analysis, the nature of
charge screening in the model differs significantly from that which appears in the large N solution.
Large N leads to a ``quark model'' view of both singlet and nonsinglet mesons, which are loosely bound but
confined $z^+z^-$ states held together by the linear Coulomb force associated with the dynamically generated
$F_{\mu\nu}^2$ term. There would thus be a constant density of topological charge (electric field) between
the two constituents of the bound state. One might then expect Euclidean topological charge distributions to 
be dominated by large coherent two-dimensional bubbles of topological charge, whose thickness was determined
by the confinement scale. Not only is this inconsistent with what is seen in the Monte Carlo, but it also
would produce a positive TC correlator for distances less than the confinement length, violating the spectral requirement.
It has been argued \cite{Rabinovici,Samuel} that the large N solution is actually
misleading in this regard because the large N saddle-point approximation entails a subtle violation of
Elitzur's theorem. This argument suggests that the nature
of the screening of the $U(1)$ charge is more accurately represented by the lattice strong coupling expansion,
where it is easily seen that, because of the absence of a bare gauge kinetic term, screening takes place 
{\it ultralocally}, in the sense that the only nonvanishing terms are those in which the net current flowing on
every link is zero. This phenomenon, which has been referred to as ``superconfinement,'' strongly suggests that,
even in the continuum limit, charge screening should take place locally. The fact that we observe essentially
one-dimensional coherent regions of topological charge, which have a typical thickness proportional to the lattice
spacing, appears to support the superconfinement view of charge screening in the $\CP$ models. Further Monte Carlo
studies of charge screening in the $\CP$ models might shed additional light on this issue.
  
The view of QCD dynamics provided by AdS/CFT holography constitutes a powerful new framework from which to explore
the structure of the QCD vacuum  \cite{Gross}. The issue of topological charge structure and $\theta$-dependence
in QCD lies at the heart of the AdS/CFT correspondence. As discussed by Witten \cite{Witten98}, the string theory
dual of QCD topological charge is Ramond-Ramond charge in IIA string theory. This is the fundamental solitonic,
or ``magnetic'' charge of the theory which is not carried by ordinary string states, but is carried by D-branes.
Comparing Witten's discussion of $\theta$-dependence from AdS/CFT holography with the earlier, purely four-dimensional
discussion of Luscher \cite{Luscher78}, it is clear that the ``Wilson bag'' 3-surface
is holographically dual to a wrapped 6-brane in Witten's description. Both of them have the defining
property that the local value of $\theta$ jumps by $2\pi$ across the surface. 

The possibility of directly confronting detailed aspects of the AdS/CFT correspondence with Monte Carlo data is 
particularly exciting. Ongoing Monte Carlo experiments on the coherent structures in both $QCD$ and $\CP$ should
provide further, more detailed, tests of the Wilson bag interpretation. 

We are grateful to P. Arnold, P. Fendley, and Y. Lian for discussions on these and related topics. This work was supported
in part by the Department of Energy under grant DE-FG02-97ER41027.
%%%%%%%%%%%%%%%%%%%%%%%%%%%%%%%%%%%%%%%%%%%%%%%%%%%%%%%%%%%%%%%%%%%%%%%%%%%

\end {document}